%% file: mmtf-6c.tex
\documentclass[12pt]{article}

\usepackage[final,pdfusetitle, pdftex]{hyperref}
\usepackage{amsmath,amssymb,amsthm}
\usepackage{enumitem}
\usepackage{authblk}
\usepackage{geometry} \usepackage[T1]{fontenc}
\usepackage[utf8]{inputenc}
\usepackage{tikz}

\usepackage{autonum} 
\usepackage{hyperref} \usepackage{verbatim}

\include{all-def}

\DeclareMathOperator{\id}{Id}

\newcounter{MODUS}
 \ifnum\theMODUS=0
 \newcommand{\DETAILS}[1]{}
\else
\newcommand{\DETAILS}[1]{{

    \medskip  \mygreen{IN DETAIL:
       $\langle\hspace{-0.3ex}\langle$#1$\rangle\hspace{-0.3ex}\rangle$}}}
 \fi 
 \newcommand{\DETS}[1]{{

     \medskip \mygreen{IN DETAIL:
       $\langle\hspace{-0.3ex}\langle$#1$\rangle\hspace{-0.3ex}\rangle$}}}

\newcommand{\DS}{\displaystyle} \DeclareMathOperator{\sign}{sign}
 
 \renewcommand{\hat}{\widehat}
\DeclareMathOperator{\arsinh}{arsinh}

\newtheorem{theorem}{Theorem}[section] %
\newtheorem{lemma}[theorem]{Lemma} 

\newtheorem{proposition}[theorem]{Proposition}
\newtheorem{remark}[theorem]{Remark} %

\begin{document}
\title{Onset of pattern formation in thin ferromagnetic films with perpendicular
  anisotropy}
\author[1]{Birger Brietzke} %
\author[2]{Hans Knüpfer}
\affil[1]{\small{ Institute of Applied Mathematics\\Heidelberg University\\Im Neuenheimer Feld 205\\
    69120 Heidelberg, Germany\\
    birger.brietzke@alumni.uni-heidelberg.de}}
\affil[2]{\small{ Institute of Applied Mathematics\\Heidelberg University\\Im Neuenheimer Feld 205\\
    69120 Heidelberg, Germany\\
    knuepfer@uni-heidelberg.de}\\
(corresponding author)}

\maketitle

 \begin{abstract}
   We consider the onset of pattern formation in an ultrathin ferromagnetic film
   of the form $\t \Ome_t := \t \Ome \times [0,t]$ for $\t \Ome \CUS \R^2$ with preferred
   perpendicular magnetization direction. The relative micromagnetic energy is
   given by
   \begin{align}
      \EE[M]\ %
      &= \ \int_{\t \Ome_t}  d^2 |\nabla M|^2+ Q \int_{\t \Ome_t} (M_1^2+M_2^2)  %
        + \int_{\R^3} |\HH(M)|^2 - \int_{\R^3} |\HH(e_3 \chi_{\t \Ome})|^2,
   \end{align} %
   describing the energy difference for a given magnetization
   $M : \R^3 \to \R^3$ with $|M| = \chi_{\t \Ome_t}$ and the uniform
   magnetization $e_3 \chi_{\t \Ome_t}$.  For $t \ll d$, we establish the
   scaling of the energy and a BV-bound in the critical regime where the base
   area of the film is of order
   $|\t \Ome| \sim (Q-1)^{1/2} d e^{\frac {2\pi d}t \sqrt{Q-1}}$. We furthermore
   investigate the onset of non--trivial pattern formation in the critical
   regime depending on the size of the rescaled film.

   \begin{center}
   \end{center}

 \end{abstract}

\tableofcontents

\section{Introduction and statement of main result}

Ferromagnetic materials are a complex class of solids which are capable of
forming a wide range of spatially ordered magnetization patterns.  These
patterns can be tuned by e.g. changing the crystaline strucure of materials or
applying external fields which has lead to a variety of applications, e.g. in
data storage (see e.g.  \cite{MoonEtal-2021}). From a mathematical point of view
these patterns can be understood as ground states of the underlying energy
functional \cite{Brown63,DeSimone2004,Hubert2008}. From the perspective of the
variational principle, the formation of magnetic domains --- large regions with
uniform magnetization --- has been studied in the physical literature
\cite{Landau-1935,Kittel-1946,WangEtal-2016}, see also \cite{Hubert2008} for an
overview as well as the mathematical literature
(e.g. \cite{CK-1998-2D,CKO-1999-3D,conti00,KM-2011-branching,otto10,
  GC-1999-magnetic,Carbou,DKMO,KohnSlastikov-2005}). In the last years there has
been increased interest in the study of extremely thin ferromagnetic films with
thickness of only a few atomic layers. In such films due to surface effects
magnetization in perpendicular direction to the film plane is energetically
preferred \cite{WadasEtal-1998AP,KronsederEtal-2015NatCom}. One feature of these
films is that they exhibit the formation of so called \textit{bubble} or
\textit{stripe domain} patterns and that the domain size grows exponentially in
terms of the inverse of the film thickness (cf. \eqref{def-s}) as experimental
and numerical observations suggest \cite{KaplanGehring-1993,Condette-2010}. In
\cite{KMN-2019}, these findings have been rigorously confirmed in a periodic
setting by minimization of energy. In this work, we consider a domain of
critical size, associated with the onset of pattern formation. In this paper, we
confirm the above scaling laws for the case of a finite film and furthermore
derive conditions for the presence and absence of non--trivial patterns,
depending on the film size. Mathematically, we have to deal with an energy
functional where local terms of diffuse perimeter type compete with the impact
of the nonlocal magnetostatic energy. In the critical regime, these two parts of
the energy have the same magnitude which makes the analysis interesting.

\medskip

Let $\t \Ome_t = \t \Ome \times [0,t]$ for a bounded domain $\t \Ome \SUS
\R^2$. In a partially non-dimensionalized form the (relative) energy associated
with the magnetization $M : \R^3 \to \R^3$ with $|M| = 1$ in $\Ome$ and $M = 0$
else, is
\begin{align}
  \EE[M]\ %
  &= \ \int_{\t \Ome_t}  d^2 |\nabla M|^2+ Q \int_{\t \Ome_t} (M_1^2+M_2^2)  %
    + \int_{\R^3} |\HH(M)|^2 - \int_{\R^3} |\HH(e_3 \chi_\Ome)|^2.
    \label{def-EE}
\end{align} %
The first three integrals above are the Landau-Lifshitz energy associated with
the magnetization \cite{Landau-1935}. In particular, the first integral is
called the \textit{exchange energy} and reflects the preference of neighboring
spins to be aligned. The second integral describes the fact that the considered
material energetically prefers a direction of the magnetization normal to the
film plane and is called \textit{anisotropy energy}. The third integral is the
\textit{magnetostatic} (or stray field) energy where the stray field operator is
the projection on rotation free vector fields, i.e.
\begin{align} \label{def-HH} %
  \HH(M) \ = \ - \nabla \Phi, \qquad\qquad \text{where $\Phi \in H^1(\R^3)$ solves
  } \ \Delta \Phi \ = \DIV M.
\end{align}
With the last integral we have subtracted the energy of the uniform
magnetization $e_3 \chi_{\t \Ome_t}$; this term does not depend on the specific
magnetization. Hence, $\EE$ is the relative energy of the magnetization $M$ with
respect to the energy of the uniform magnetization. Finally, the parameter $d$
is called exchange length and the dimensionless parameter $Q > 0$ is called
quality factor. We consider thin films where the thickness is much smaller than
the exchange length, i.e.  $t \ll d$. The quality factor is fixed throughout
this work and we assume $Q>1$ which corresponds to a so called hard material.

\medskip

A good way to understand the magnetostatic energy is from the perspective of
electrostatics: In fact, in view of \eqref{def-HH}, the magnetostatic field is
created by the distributional divergence $\DIV M$ in the same way as the
electrostatic field is created by charge density in electrostatics. With this
analogy in mind, $\DIV M$ is hence denoted as \textit{magnetic charge}
(density). In turn $\DIV M$ consists of two parts: The absolutely continuous
part $(\DIV M)_{ac}$ are the so called \textit{volume charges}; any non--zero
normal component $(\DIV M)_{s} = M \cdot n \HH^2_{|\p \t \Ome_t}$ at the
boundary $\p \t \Ome_t$ with outer normal $n$ are the so called \textit{surface
  charges}.

\medskip

\begin{figure}
  \centering %
  \includegraphics[height=4cm]{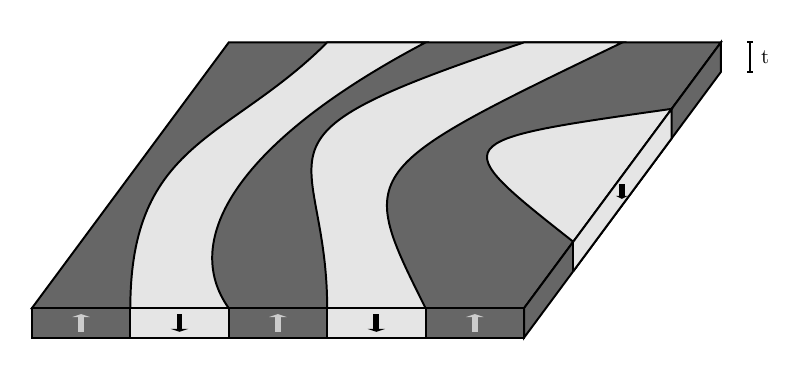} \qquad %
  \caption{Sketch of a stripe domain pattern structure in a thin ferromagnetic
    film.}
  \label{fig-domains}
\end{figure}

The analysis of this functional is non-trivial as it amounts to a nonconvex and
nonlocal variational problem. The local parts of the energy in \eqref{def-EE},
i.e. the exchange energy and anisotropy energy, favor a uniform magnetization,
at the same time a uniform magnetization leads to a high magnetostatic
energy. The competition of the local and nonlocal (magnetostatic) part of the
energy hence leads to the formation of magnetostatic domains, i.e. extended
regions of uniform magnetization which are separated by thin transition layers
where the magnetization rotates rapidly (see
Fig. \ref{fig-domains}). Experimental and numerical findings
\cite{KaplanGehring-1993, Condette-2010} indicate the following scaling laws for
the typical domain width and the ground state energy:
\begin{align}  %
  s \ %
  &:= \ d \ \frac{e^{\frac {2\pi d}t \sqrt{Q-1}}}{\sqrt{Q-1}} \quad
  &&\text{scaling of typical domain width}, \label{def-s} \\ \qquad %
  \inf_M \EE[M] \ %
  &\sim \ d t^2 \ \frac{e^{\frac {2\pi d}t \sqrt{Q-1}}}{\sqrt{Q-1}} \ %
  &&\text{scaling of ground state energy.} \ \label{def-e} 
\end{align}
These scaling laws have been investigated in \cite{KMN-2019} in a periodic
setting. In this paper, the authors correspondingly identified a subcritical and
a supercritical regime. In particular, they showed that the above scaling laws
holds in the supercritical regime. In this work we investigate the critical
regime for a finite magnetic sample associated with the onset of formation of
magnetic domains and the corresponding ground state energy. We confirm these
scaling laws in the critical regime in the case of a finite sample. Due to the
nonlocal character of the energy, this requires some adaption of the methods
used in \cite{KMN-2019}.  While the analysis in \cite{KMN-2019} is mainly
concerned with the situation of the subcritical and supercritical regime, in
this paper we precisely consider the critical regime at the onset of pattern
formation and we use a rescaling of the model which is suited to the critical
regime. We then show that the onset of non--trivial pattern formation occurs in
the critical regime and give some conditions for the presence and absence of
non--trivial pattern formation.  We note that, in general, the parameters in
\eqref{def-EE} also dependent on the temperature which hence also impacts the
domain structure \cite{Yamanouchi-2011MagLet}. Other effects such as the
Dzyaloshinskii-Moriya interaction are also relevant in some parameter regimes
and have been considered in the literature as well (see e.g.
\cite{MeierEtal-2017PRB,Lemesh-2017PRB}).

\medskip

In view of the above considerations we rescale the energy according to the
scaling of the ground state energy and the minimum of the length scale according
to the typical width $s$ of the magnetic domains or the sample, i.e.
\begin{align} \label{full-E} %
  E_{\eps}[M] \ &:= \ \frac {\pi}{st^2} \EE[\t M] \qquad\qquad %
                  \text{where $M(x,x_3) \ := \ \t M(s x,t x_3) $} %
\end{align}
for $(x,x_3) \in \R^2 \times \R$ and where we introduce the non--dimensional
parameter
\begin{align} \label{def-eps} %
  \eps \ := \  \ e^{- \frac {2\pi d}t \sqrt{Q-1}}, \qquad 
\end{align}
which represents the ratio between the typical width of transition layers versus
the typical domain width. Our problem can hence be characterized by the two
dimensionless parameters $Q$ and $\eps$. The thin film regime $t \ll d$
corresponds to the regime $\eps \ll 1$. We will assume $Q > 1$ throughout the
paper. With the notation $\Ome_1 := \Ome \times [0,1] \SUS \R^3$, we
correspondingly investigate the energy $E_\eps$ on the class of magnetizations
\begin{align}
  \AA \ = \ \Big \{ M \in L^2(\R^3:\R^3) \ 
  : \ \text{$M \in H^1(\Ome_1)$, $|M| = 1$ in $\Ome_1$ and $M = 0$ else}  \Big \}.
\end{align}

\medskip

Variation of the magnetization in thickness direction is energetically
unfavorable for thin films and penalized in terms of the exchange part of the
energy.  We hence reduce this problem in Section \ref{sec-dimred} to a
two--dimensional model, formulated in terms of the average magnetization over
the film height, given by
\begin{align} \label{def-avg} 
  \OL M(x) \ := \ \Big(\dashint_0^t M(s x, x_3) dz \Big) \chi_{[0,t]}(x_3). \qquad\qquad 
\end{align}
We first give a criterion on the onset of pattern formation in terms of the size
of the underlying domain $\Ome$:
\begin{theorem}[Onset of domain formation] \label{thm-critical1}
  \text{} Let $\Ome \CUS \R^2$ be convex and bounded. Then there is a universal
  constant $\eps_0 > 0$ such that for $\eps \in (0,\eps_0)$ we have
  \begin{enumerate}
  \item (Domains with small diameter) If $\diam \Ome \ %
    < \ \frac {\pi}{2e} (1 - \frac 2{|\ln \eps|})$, then
    \begin{align} \label{only-2} %
      \inf_{M \in \AA, M = \OL M} E_\eps[M] \   = \ 0. %
    \end{align}
    Furthermore, the only two minimizers in \eqref{only-2} are given by
    $M = \pm e_3 \chi_{\Ome_1}$.
  \item\label{F-it-Eneg} (Minimal energy for large domains) There are universal
    constants $0 < c, R < \infty$ such that for
    $\Ome^{(R)} := \{ x \in \Ome : \dist(x,\Ome^c) \geq R \} \SUS \Ome$ we have
    \begin{align} \label{F-thelob-0} %
      - \frac{\pi^2 e }4 |\Ome| \ \leq \ \inf_{M \in \AA} E_\eps[M] \ \leq \ - c
      |\Ome^{(R)}|.
  \end{align}
  \end{enumerate}
\end{theorem}
Theorem \ref{thm-critical1} characterizes the onset of magnetic domains. In
particular, statement \textit{(i)} shows that for samples with small diameter
and for magnetizations with $M = \OL M$, the energy is minimized by
$m = \pm e_3 \chi_{\Ome_1}$. For general magnetizations a related estimate holds
true up to a small error as stated in Lemma \ref{lem-small}. For larger films,
statement \ref{F-it-Eneg} shows that the uniform magnetization is not optimal
anymore. Furthermore, this statement confirms the scaling law \eqref{def-e} for
the ground state energy, formulated in the rescaled variables.

\medskip

The theorem is concerned with the scaling law \eqref{def-s} for the average size
of the domain.  These statements are formulated in terms of the total
interfacial length which can be seen as a measure for the typical width of the
domains (since $\Ome$ has area of order one). In turn, this (scalingwise)
implies that the average width of the magnetic domains is of order one. We have
the following rigorous justification for these observations:
\begin{theorem}[BV bounds and compactness] \label{thm-critical2} \text{} With
  the assumptions of Theorem \ref{thm-critical1} and with $\OL M$ defined in
  \eqref{def-avg} we have
  \begin{enumerate}
\item\label{it-wallhigh} (BV estimate) For the average magnetization
  \eqref{def-avg} we have
    \begin{align} 
      \Big|\ln \Big(\frac{\NNN{\nabla \OL M_3}{\Ome_1}}{\pi^2 e^2|\Ome|} \Big) \Big| \NNN{\nabla \OL M_3}{\Ome_1}   \ 
      &\leq \ E_\eps[\OL M] + \frac{\pi^2 e }4 |\Ome| \qquad\qquad
        \FA{M \in \AA.}
        \label{X-F-est}
    \end{align}
  \item\label{it-wallest} %
    (BV estimate for low energy configurations) %
    For any $\alp > 0$ there are constants $c_\alp, C_\alp > 0$ such that for
    any $M \in \AA$ with $E_\eps[M] < - \alp |\Ome|$ we have
    \begin{align}
      c_\alp |\Ome| \ \leq \ \NNN{\nabla \OL M_3}{\Ome_1} \ \ %
      \leq \ C_\alp |\Ome|.
    \end{align}
  \item\label{it-compact} %
    (Compactness) For any family $M_\eps \in \AA$ with
    \begin{align}
      \limsup_{\eps \to 0} \ E_\eps[M_\eps] \ 
      < \ \infty
    \end{align}
    there is $m \in BV(\R^2, \pm e_1)$ with $|M| = 1$ in $\Ome$ and $M = 0$ else
    such that for a subsequence $\eps \to 0$ and with
    $(m \otimes \chi_{[0,1]})(x,x_3) := m(x)\chi_{[0,1]}(x_3)$ we have
    \begin{align}
      M_\eps \ \to \ m \otimes \chi_{[0,1]} \qquad\qquad \text{in $L^1(\Ome_1)$.}
    \end{align}
  \end{enumerate}
\end{theorem}
The above theorem gives a rigorous justification for the conjecture on the
scaling law \eqref{def-s} for the typical domain size. Statement
\ref{it-wallhigh} shows that a sharp bound for the BV norm in terms of the
nonlocal energy.  In particular, for configurations with small relative energy
\ref{it-wallest} shows that the total length of the interfaces is of order $1$.
Finally, \ref{it-compact} gives a corresponding compactness result which is a
direct consequence of these bounds on the BV norm.

\medskip

In the course of the proof, we first show that the energy \eqref{full-E} can be
approximated in our regime by another two--dimensional energy, given by
\begin{align} \label{def-F} 
  F_\eps[m] &= |\ln \eps| \int_{\Ome} \Big(\frac{\eps}2 |\nabla m|^2+\frac
              1{2\eps} (1-m_3^2) \Big) dx - \frac 18 \iint_{\R^4} \frac {|m_3(x)
              - m_3(x')|^2}{|x-x'|^3} dx dx'.
\end{align}
In fact, the corresponding results to Theorems \ref{thm-critical1} and
\ref{thm-critical2} also hold for the energy $F_\eps$. The natural domain
for the energy $F_\eps$ is the class of magnetizations
\begin{align}
  \OL {\AA} \ 
  := \ \big \{ m \in L^2(\R^2;\R^2) \ 
  : \ \text{$m \in H^1(\Ome)$, $|m| = 1$ in $\Ome$ and $m = 0$ else}  \big \}.
\end{align}

Without the nonlocal term, by standard Modica--Mortola theory the first two
terms on the right hand side of \eqref{def-F} would $\Gam$--converge to the BV
norm of $m_3$. The second term is nonlocal and is the homogeneous
$H^{\frac 12}$--norm of $m_3$. Note that the BV--norm has the same scaling as
the $H^{\frac 12}$--norm which is a major difficulty in the proof of the results
and which explains the appearance of logarithms in the statement of Theorem
\ref{thm-critical2}.  In the proof we will show that the leading order
contributions of these two terms cancel in the limit $\eps \to 0$ and the bound
on the ground state energy is derived from the remaining term.

\medskip

We note that the corresponding model in a periodic setting has been investigated
in \cite{KMN-2019} by Muratov, Nolte and the second author. The authors
identified a subcritical and a supercritical regime. In the subcritical regime,
they then show a $\Gam$--limit result where the energy converges to a local
limit energy. In the supercritical case, energy bounds are derived which confirm
the scaling laws \eqref{def-s}--\eqref{def-e}.  In this work we consider the
critical regime in the geometry of a finite sample where additional surface
charges may occur at the lateral boundary of the magnetic film. In this setting,
we show that the same scaling laws still hold. Furthermore, we derive new
conditions on the onset of pattern formation within the critical regime.

\medskip

We also note that a corresponding sharp interface versions of the energy
\eqref{def-EE} has been considered by different authors. In these models,
compared to the energy \eqref{def-F}, the local part of the energy is replaced
by the perimeter term. At the same time, the $H^{\frac 12}$--norm is replaced by
a regularized version of this norm with an $\eps$--dependent high frequency
cut--off. This is reminiscent of the results of Bourgain, Brezis, Mironescu
\cite{BBM01} and D\'avila \cite{Da02} which show the convergence of a class of
nonlocal perimeter functionals towards the $BV$--norm. A related
two--dimensional model energy has been studied by Muratov and Simon in
\cite{MS19}, where the authors show among other results a $\Gam$-convergence
result. For a different approximation of the nonlocal term, a corresponding
$\Gam$-convergence result has been given by Cesaroni and Novaga
\cite{CesaroniNovaga-2020}. A corresponding $\Gam$--convergence result in
arbitrary dimensions and for a more general class of approximations of the
nonlocal energy has been shown in \cite{KnuepferShi-preprint} by Shi together
with the second author. These results can be seen as second order extensions of
the result in \cite{Da02}.

\medskip

More generally, pattern formation has also been investigated in related models
which include a competition between a local isoperimetric term and an opposing
nonlocal term such as e.g.  the Ohta-Kawasaki energy
\cite{AceFusMor,AlbertiChoksiOtto-2009,ChoksiPeletier-2010,ChoksiPeletier-2011,GolMurSerI},
liquid drop models \cite{CicSpa,JulinPisante-2017,MorSte} or elastic materials
(these references are certainly not complete). The methods used are typically
specific for the model at hand. In particular, in the above models the nonlocal
term does not have the same scaling as the perimeter contrary to the problem
considered here.

\medskip

\textbf{Structure of paper:} In Section \ref{sec-dimred} we derive estimates on
the stray field which show that the energy \eqref{full-E} can be approximated by
the two--dimensional energy \eqref{def-F}. The proofs of Theorem
\ref{thm-critical1} and Theorem \ref{thm-critical2} are then given in Sections
\ref{sec-lower} and \ref{sec-upper}. In particular, the theorems are a
consequence of Lemma \ref{lem-small} and Propositions \ref{prp-lowfull},
\ref{prp-compact} and \ref{prp-constr}.

\medskip

\textbf{Notation:} We write $A \lesssim B$ if $A \leq C B$ for
some universal constant $C < \infty$.  The notations $\gtrsim$ and $\sim$ are
defined analogously.

\medskip

For two sets $E,F \SUS \Rn$ we write $E \CUS F$ if the set $E$ is compactly
embedded in the set $F$. By $B_\d(E) := \{ x \in \Rn : \dist(x,E) < \d \}$ we
denote the $\d$-- neighborhood of the set $E$.  For a set $\Ome \SUS \Rn$ with
non--zero interior we write $|\Ome| := \LL^n(\Ome)$ and
$|\p \Ome| := \LL^{n-1}(\p \Ome)$. By $\chi_\Ome$ we denote the characteristic
function of the set $\Ome$. The total variation of a function on the set $\Ome$
is written as $\NNN{\nabla f}{\Ome} := \int_\Ome |\nabla f|$.

\medskip

We denote points in $\R^3$ usually by $\OL x = (x,x_3) \in \R^3$. On the other
hand, the magnetization vector field $M : \R^3 \to \R^3$ is written in the form
$M = (M',M_3)$. We use the notation $M := m \otimes \chi_{[0,t]}$ for the
function $M(x,x_3) = m(x)\chi_{[0,t]}(x_3)$. For the two--dimensional Fourier
transform we use the convention
\begin{align}
  \hat f(\xi) \ 
  = \ \frac 1{2\pi} \int_{\R^2} f(x) e^{-i\xi \cdot x} \ dx,
\end{align}
so that in particular Plancherel's identity holds with prefactor $1$.

\section{Dimensional reduction} \label{sec-dimred} %

In this section, we show that the two--dimensional energy \eqref{def-F} is a
good approximation for the initial energy in the considered regime of
thin--films. Due to the specific geometry of our sample, it is useful to
differentiate between directions within the film plane and normal directions. We
hence consistently write $(x,x_3) \in \R^2 \times \R$ for points in $\R^3$. We
first recall that the stray field operator \eqref{def-HH} is the orthogonal
projection onto gradient fields:
\begin{lemma}[Stray field] \label{lem-stray-1} 
  For any $M \in H^1(\R^3; \R^3)$ we have
  \begin{align} \label{id-HH} %
    \HH(M)(x,x_3) \ = \ \nabla \int_{\R^3} K(x-x',x_3-x_3') \DIV M(x',x_3') \
    dx' dx_3',
     \end{align}
     where the kernel $K : \R^2 \times \R \to \R$ and its Fourier transform (in
     terms of $x$) are
     \begin{align} \label{fund-sol} 
       K(x,x_3) \ := \ \frac 1{4\pi\sqrt{|x|^2 + x_3^2}}, \qquad \hat K (\xi,x_3) \ =
       \ \frac 1{4\pi|\xi|} e^{-|x_3| |\xi|}.
   \end{align}
   Furthermore, $\HH : L^2(\R^3; \R^3) \to L^2(\R^3; \R^3)$ is linear,
   $\HH^2(M) = \HH(M)$ and
     \begin{enumerate}
     \item 
       $\DS \int_{\R^3} \HH[M] \cdot \big( \HH[\t M] - \t M \big)  \ = \ 0$
       \qquad\qquad 
       $\forall \t M \in H^1(\R^3; \R^3).$
     \item 
       $\DS \int_{\R^3} |\HH[M]|^2  \ $
       $\DS \ \leq \ \int_{\R^3} |M|^2$.
     \end{enumerate}
   \end{lemma}
   \begin{proof}
     The kernel $K$ is just the Newton kernel in $\R^3$ which justifies
     \eqref{id-HH}. A standard integral formula \cite[eq
     6.554.1]{Zwillinger-Book} then yields
  \begin{align}
    \hat K(\xi) \ 
    = \ \frac 1{2\pi} \int_{\R^2} \frac {e^{-i\xi \cdot x}}{4\pi \sqrt{|x|^2 + x_3^2}} \ dx \ 
    = \ \int_0^\infty \frac {J_0(|\xi| r)}{4\pi \sqrt{r^2 + x_3^2}}  \ r dr \ 
    = \ \frac 1{4\pi} \frac 1{|\xi|} e^{-|x_3| |\xi|},
  \end{align}
  where $J_0$ is the Bessel function of first kind.  We next give the proof for
  (i)--(ii): Multiple integration by parts yields
\begin{align}
  \int_{\R^3} \HH(M) \cdot \HH(\t M) \ 
  = \   - \int_{\R^3} \phi \cdot \Delta \t \phi \ 
  = \   \int_{\R^3}  \phi \cdot \DIV \t M  \ 
  = \   \int_{\R^3} \HH(M) \cdot \t M.
\end{align}
Estimate (ii) follows from (i) and an application of Cauchy-Schwarz.
\end{proof}
\DETAILS{We note that $\hat K$ can alternatively be derived directly from the equation
$\Delta \phi = \DIV M$ for the magnetostatic potential (cf. \cite{KMN-2019}). In
terms of the Fourier transform we have
$\p_3^2\hat{\phi}(\xi,x_3)-|\xi|^2\hat{\phi}(\xi,x_3) = \hat{\DIV m}$.  Then
$\hat K$ is the fundamental solution of this ODE, i.e.
$\p_3^2 \hat K(\xi,x_3)-|\xi|^2 \hat K(\xi,x_3)\ =\ \delta_0(x_3)$. Here, we need to
have an additional factor $2\pi$ from the convolution identity for the Fourier
transform.

\medskip

}Variation of the magnetization in thickness direction is energetically
unfavorable for thin films and penalized in terms of the exchange part of the
energy.  For $M \in H^1(\Ome_{t};\R^3)$ we hence consider the corresponding
magnetization \eqref{def-avg} which is averaged in the thickness direction.

\medskip

The following result is a version of \cite[Theorem 5.2]{KMN-2019} for finite
domains. The different geometry requires some adaptions to the proof since the
magnetization might be discontinous at the lateral boundary. This is reflected
on the right hand side of \eqref{md-claim} which incorporates the lateral sample
boundary $\p \Ome \times [0,t]$:
\begin{theorem}[Reduction of stray field]\label{thm-strayred}
  Let $\t \Ome \CUS \R^2$ and let $\t \Ome_t := \t \Ome \times [0,t]$. Let
  $M = (M',M_3) \in \AA$ and let $\OL M$ be given by \eqref{def-avg}. Then
  \begin{align} \label{md-claim} %
    \int_{ \R^3} |\HH[M]|^2  \ &= \ \int_{ \R^3} |\HH[\OL{M}]|^2 + \OO \Big(
                                     t |B_{t}(\p \t \Ome) \cap \t \Ome| + t^2
                                     \int_{\t \Ome_t}|\nabla M|^2 \Big).
  \end{align}
\end{theorem}
\begin{proof}
  We define $U := M-\OL{M}$.  Since $\OL U = 0$ by Lemma \ref{lem-stray-1} and
  Poincar\'e's inequality we have
  \begin{align} \label{uest} 
    &\int_{\R^3} |\HH(U)|^2 \ \leq \int_{\t \Ome_t} |U|^2  \ %
      \leq  \ t^2 \int_{\t \Ome_t} |\p_3 M|^2 
      = \ \OO \Big( t^2 \int_{\t \Ome_t}|\nabla M|^2 \Big).
  \end{align}
  We write $M = \OL M + U$ and expand the expression on the left hand side of
  \eqref{md-claim} using the fact that $\HH$ is a linear operator. By
  \eqref{uest} it is then enough to show that
  \begin{align}
    \bigg|\int_{\t \Ome_t} \HH(\OL M) \cdot U  \bigg| \ 
    = \ \OO \Big( t |B_{t}(\p \t \Ome) \cap \t \Ome|  + t^2 \int_{\t \Ome_t}|\nabla M|^2 \Big). \label{tmt} 
  \end{align}
  For $\d > 0$, we choose a cut-off function $\zet_\d \in H^1(\R^2)$ by
  $\zet_\d(x)$ $:=$ $\min$ $\{ 1, \frac 1{\d} \dist(x,\p \t \Ome) \}$ for
  $x \in \t \Ome$ and $\zet_\d = 0$ else.  Then we have $0 \leq \zet_\d \leq 1$
  and $\zet_\d = 0$ in $\t \Ome^c$ and $\zet = 1$ in $\t \Ome \BS \t B$ where
  $\t B := B_{\d}(\p \t\Ome) \cap \t\Ome$. We now define
  $M_\d(x,x_3) \ := \ \t \zet_\d(x) \OL M(x,x_3)$.  By Lemma \ref{lem-stray-1},
  by Jensen's inequality and with the choice $\d := t$ we then have
  \begin{align}
    \int_{\t \Ome_t} |\HH(M_\d) - \HH(\OL M)|^2 \ 
    &\leq \ \int_{\t \Ome_t} |M_\d - \OL M|^2 \ \leq \ \d |\t B| \ = \ t |\t B|, \label{MdM-1} \\
    t^2 \int_{\t \Ome_t} |\nabla M_\d|^2  \ 
    &\leq \ t^2  \Big( 
      \frac {t|\t B|}{\d^2} + \int_{\t \Ome_t} |\nabla \OL M|^2   \Big) \ 
      = \ \OO\Big(t |\t B| + t^2 \int_{\t \Ome_t} |\nabla M|^2 \Big).
  \end{align}
  By an application of Cauchy--Schwarz and in view of \eqref{uest}, we then also have
  \begin{align}
    \bigg|\int_{\t \Ome_t} \HH(M_\d - \OL M) \cdot U  \bigg| \ 
    &= \ \OO \Big( t |\t B| + t^2 \int_{\t \Ome_t} |\nabla M|^2 \Big).
  \end{align}
  In view of these calculations, \eqref{tmt} then follows if we can show that
  \begin{align}
    X \ 
    := \ \bigg|\int_{\t \Ome_t} \HH(\OL M_\d) \cdot U  \bigg| \ 
    = \ \OO \Big( t^2 \int_{\t \Ome_t}|\nabla M|^2 \Big). \label{tmt2} 
  \end{align}
  The advantage of the new function $M_\d$ is that $M_\d = 0$ on
  $\p \t \Ome \times [0,t]$ which implies
  \begin{align} \label{adv} %
    \int_{\t \Ome_t} |\nabla' \OL M|^2 \ dx dx_3  \ %
    = \ \int_{\R^3} |\nabla' \OL M|^2 dx dx_3 \ %
    = \ \int_{\R^3} |\xi|^2 |\hat{\OL M}|^2 \ d\xi dx_3.
  \end{align}
  We use the stray field formula from Lemma \ref{lem-stray-1} and apply the
  Fourier transform in $x$. Integrating by parts and since
  $\OL M(x,x_3) = \OL M(x) \chi_{[0,t]}(x_3)$ with $P := \hat{M_\d(\xi)}$ we get
  \begin{align} \label{HH-reps} %
    \begin{aligned}
      \hat {\HH'}(\OL M)(\xi,x_3) %
      &= 4\pi \Big( \int_0^t  |\xi|^2 \hat{K}(\xi,x_3-x_3') dx_3' P' 
      - \int_0^t  i \xi \p_3  \hat{K}(\xi,x_3-x_3') dx_3' P_3 \Big) \\
      \hat {\HH_3}(\OL M)(\xi,x_3)  %
      &= 4\pi \Big(\int_0^t  \p_3^2  \hat{K}(\xi,x_3-x_3') dx_3' P_3 - \int_0^t  i \xi \p_3 \hat{K}(\xi,x_3-x_3') dx_3'  P'  
      \Big).
    \end{aligned}
    \end{align}
    We note that the antiderivative and derivative of
    $K = \frac 1{4\pi |\xi|} e^{-|\xi||x_3|}$ in $x_3$ are given by $\int K$ $=$
    $\frac{\sign x_3}{4\pi|\xi|^2} ( 1 - e^{-|\xi| |x_3|}) + c$ and $\p_3 K$ $=$
    $\frac {\sign x_3}{4\pi} e^{-|\xi| |x_3|}$.  With these identities we
    compute the Fourier multipliers in \eqref{HH-reps} for $x_3 \in (0,t)$. We
    get
  \begin{align}
    \mu_1(\xi,x_3) \ &:= \ 4\pi |\xi|^2 \int_0^t   \hat{K}(\xi,x_3-x_3') \ dx_3'   \ 
    = \ \frac 12 \big( e^{-|\xi| |x_3-t|}  + e^{-|\xi| |x_3|} - 2 \big), \\
    \mu_2(\xi,x_3) \ &:= \  4\pi  |\xi| \int_0^t \p_3 \hat K(\xi,x_3-x_3')  \ dx_3'  \ 
    = \ \frac 12 \big( e^{-|\xi| (t-x_3)} - e^{-|\xi| x_3} \big), \\
    \mu_3(\xi,x_3) \ &:= \  4\pi  \int_0^t \p_3^2 \hat K (\xi,x_3-x_3')  \ dx_3'   \ 
    = \  \frac 12 \big( e^{-|\xi| (t-x_3)} + e^{-|\xi| x_3} \big).
  \end{align}
  We now use the fact that $U$ has a vanishing average, i.e. $\OL U = 0$, to
  conclude that
  \begin{align} \label{U-null-1} 
    X \ 
    &= \ \Big|\int_{\R^3} \HH(M_\d) \cdot U \ dx \Big| \ 
      = \ \Big|\int_{\R^3} (\HH(M_\d) - M_{\d,3} e_3) \cdot U \ dx \Big|.
  \end{align}
  Hence, by Plancherel's identity we can estimate $X$ in the form
  \begin{align} \label{U-null} 
    X \ 
    &= \ \Big|\int_{\R^3} (\HH(M_\d)  - M_{\d,3} e_3) \cdot U \ dx \Big| \\
    &\upref{HH-reps}\leq  \ \sum_{i=1}^3 \int_{\R^3}  |\mu_i - \d_{i3}| |\hat{M_\d}(\xi)| |\hat U(\xi,x_3)| \ d\xi dx_3.
  \end{align}
  By Taylor expansion we have $|\mu_i(\xi,x_3) - \d_{i3}|$ $\leq |\xi| t$ for
  $i=1,2,3$ and $|x_3| \leq t$. Also using Cauchy--Schwarz and \eqref{uest} and in
  view of \eqref{adv} the proof is concluded by
    \begin{align}
      X \ 
      &\upref{U-null}\leq \ t \int_0^t \int_{\R^2}  |\xi| |\hat{M_\d}(\xi)| |\hat U(\xi,x_3)| \ d\xi dx_3  \Big| \\
      &\leq \ t \Big( \int_{\t \Ome} |\nabla' M_\d|^2  \ dx dx_3 \Big)^{\frac 12} \Big( \int_{\R^3} |U|^2  \ dx dx_3 \Big)^{\frac 12} \ 
        \lesssim \ t^2 \int_{\t \Ome} |\nabla M_\d|^2  \ dx.
  \end{align}
\end{proof}
  We next derive the kernels for the main contribution of
  the stray field in thin--films. 
\begin{theorem}[Representation of stray field]\label{thm-stray-rep}
  Let $M = (M',M_3) \in H^1(\R^3;\R^3)$ with $\spt M \SUS \R^2 \times [0,t]$ and
  let $\OL M$ be given by \eqref{def-avg}.  Then
  \begin{align} \label{ortho} 
     \DS \int_{ \R^3} |\HH[\OL M]|^2 \ 
    = \ \int_{ \R^3} |\HH[\OL{M_3}e_3]|^2 + \int_{ \R^3} |\HH[\OL{M'}]|^2 \ dx.
  \end{align}
  Furthermore, we have the representations:
  \begin{align} \label{it-stray-vert} 
    \int_{\R^3} |\HH[\OL{M_3}e_3]|^2 \ dx 
    &=\ t \int_{\R^2} {\OL M_3}^2 \ dx - \frac {t^2}{8\pi} \iint_{\R^4} \Gam(\tfrac 1t
      |x-x'|) \frac{|{\OL M_3}(x) - {\OL M_3}(x')|^2}{|x-x'|^3}
      \ dx dx' \\
    \int_{\R^3} |\HH[\OL{M'}]|^2 \ dx 
    &=\ \frac {t^2}{4\pi} \iint_{\R^4} \Theta(\tfrac 1t |x-x'|) \frac{\DIV \OL M'(x)
      \DIV \OL M'(x')}{|x-x'|} \ dx dx' \label{it-stray-tang} 
  \end{align}
  where the monotonically increasing functions $\Gam$,
  $\Theta : [0,\infty) \to \R$ are given by
  \begin{align} \label{def-Gam} 
    \Gam(\alp) \ := \ 2 \alp^2 \Big(1 - \frac \alp{\sqrt{1+ \alp^2}} \Big), \quad
    \Theta(\alp) \ := \ 2 \Big(\alp \arsinh(\frac 1\alp) + \alp^2 - \sqrt{1 + \alp^2}  \Big).
  \end{align}
  In particular, $\Gam(0) = \Theta(0) = 0$ and $\Gam(\alp), \Theta(\alp) \to 1$
  for $\alp \to \infty$.
\end{theorem}
\begin{proof}
  To prove the assertion \eqref{ortho} we write $\OL M = \OL{M_3}e_3 + \OL{M'}$
  and expand the expression $\HH(\OL{M_3}e_3 + \OL{M'})$. The assertion then
  follows from
  \begin{align} \label{sym-null} 
    \int_{\R^3} \HH(\OL {M_3} e_3) \cdot \HH(\OL M') \ = \ 0.
  \end{align}
  The identity \eqref{sym-null} holds in view of the following symmetries: The
  (surface) charge density $\p_3 \OL {M_3}$ is antisymmetric w.r.t. the plane
  $\{ x_3 = \frac 12 \}$, while the charge density $\nabla' M'$ is symmetric
  w.r.t. the same plane. These symmetry properties carry over to the created
  stray fields.

  \medskip
  
  The proof of the representation formulas is based on the real space
  representation \eqref{eq-Hfourier} of the stray field energy in Lemma
  \ref{lem-stray-1}. In particular, we have
  \DETAILS{The next formula follows from
    \begin{align}
    \int_{\R^3}|\HH[M]|^2\ dx \ 
    &=\ \frac 1{2\pi} \int_{\R^6} \nabla_{x,x_3} \Phi(x-x',x_3-x_3')  \HH(M)(x')   \  dx_3 dx_3' dx dx' \\
    &=\ - \frac 1{2\pi} \int_{\R^6}  \Phi(x-x',x_3-x_3')  \nabla_{x,x_3} \cdot \HH(M)(x')   \  dx_3 dx_3' dx dx'. 
    &=\ - \frac 1{2\pi} \int_{\R^6}  \Phi(x-x',x_3-x_3')  \nabla_{x,x_3} \cdot M(x')   \  dx_3 dx_3' dx dx'.
   \end{align}
  }
  \begin{align} \label{eq-Hfourier} 
    \int_{\R^3}|\HH[\OL M]|^2\ dx \ 
    &=\ \iint_{\R^6} K(x-x',x_3-x_3')  \DIV \OL M(x) \DIV \OL M(x')   \  dx_3 dx_3' dx dx'.
  \end{align}
  We apply this formula for $M_3 e_3$ as well as for $M'$.  Integrating by parts
  two times in $x_3$ in the formula \eqref{eq-Hfourier} yields
  \begin{align}
    \int_{\R^3} |\HH[\OL{M_3}e_3]|^2 \ dx
    &=\ \iint_{\R^4} \Big( \iint_{[0,t]^2} \p_3^2 K(x-x',x_3-x_3') dx_3 dx_3' \Big) {\OL M_3}(x)  {\OL M_3}(x')   \  dx dx' \\
    &=:\ \iint_{\R^4} G_t(x-x') {\OL M_3}(x)  {\OL M_3}(x')   \  dx dx', \label{rhsof}
  \end{align}
  where
  \begin{align}
    G_t(x) \ 
     &= \ \int_0^t \Big[ \p_3 K(x,x_3-x_3') \Big]_{x_3=0}^{x_3=t} \ dx_3' \ 
      = \ 2 \big( K(x,0) - K(x,t) \big) \\
    &= \ \frac 1{2\pi|x|} \Big( 1 - \frac 1{\sqrt{1+ (\frac t{|x|})^2}} \Big) \ 
      \lupref{def-Gam}= \ \frac {t^2}{4\pi|x|^3} \Gam(\frac {|x|}t).
  \end{align}
  We write the right hand side of \eqref{rhsof} as a difference operator, i.e.
  \begin{align}
    \int_{\R^3} |\HH[\OL{M_3}e_3]|^2 dx \ 
    &= \ \iint_{\R^4} \Big( G_t(x-x') |{\OL M_3}(x)|^2  - \frac 12 G_t(x-x') |{\OL M_3}(x) - {\OL M_3}(x')|^2 \Big).
  \end{align}
  The proof of the first assertion is concluded by an application of Fubini
  together with
   \begin{align} \label{int-Gt}
     \int_{\R^2} G_t(x_3) \ dx_3 \ 
     &= \ \int_0^\infty 1 - \frac{\rho}{\sqrt{t^2 + \rho^2}} \ d\rho \
       = \ \Big[\rho - \sqrt{t^2 + \rho^2} \Big]_0^\infty \ 
       = \ t.
   \end{align}
   For the proof of the second assertion we note that for
   $\Psi(x,x_3) := \frac {|x|}{4\pi} [(1 + \frac{x_3^2}{|x|^2})^{1/2} -
   \frac{x_3}{|x|}\arsinh(\frac {x_3}x)]$, we have
   $\p_3 \Psi(x,x_3) = - \frac 1{4\pi}\arsinh(\frac {x_3}{|x|})$ and
   $\p_3^2 \Psi(x,x_3) = K(x,x_3)$. Hence, as before we calculate
    \begin{align}
      \hspace{6ex} & \hspace{-6ex} %
                     \iint_{[0,t]^2} K(x,x_3-x_3') \ dx_3dx_3' \ 
      = \ 2 \Big[ \Psi(x,0) - \Psi(x,t) \Big] \\ 
      &= \ \frac {|x|}{2\pi} \Big[\frac {t}{|x|} \arsinh(\frac {t}{|x|}) + 1 - \sqrt{1 + \frac {t^2}{|x|^2}}  \Big] \ %
      = \ \frac {t^2}{4\pi |x|} \Theta\Big(\frac {|x|}{t}\Big).
    \end{align}
    A straightforward calculation shows that both $\Gam$ and $\Theta$ are
    monotonically increasing with $\Gam(0) = \Theta(0)= 0$ and
    $\Gam(s), \Theta(s) \to 1$ for $s \to \infty$ .
 \end{proof}
 It is useful to keep the following relations between our parameters in mind:
 \begin{remark}[Parameter relations] \label{rem-parel} 
   We collect the relations of the three dimensional parameters $d$, $t$ and $s$
   in terms of the dimensionless parameters $\eps = \frac{d}{s \sqrt{Q-1}}$ and
   $Q$ where we consider the regime $\eps \ll 1$ with fixed $Q > 1$.  By
   \eqref{def-eps} we have $|\ln \eps|$ $=$ $\frac {2\pi d}{t} \sqrt{Q-1}$.  By
   \eqref{def-s}--\eqref{def-eps} we get $\frac ds = \eps \sqrt{Q-1}$, Hence,
   \begin{align} \label{le-sc} 
     \frac ds \ = \ \eps \sqrt{Q-1}, \qquad 
     \frac st \ = \ \frac{|\ln \eps|}{2\pi (Q-1)\eps}, \qquad 
     \frac dt \ = \ \frac{|\ln \eps|}{2\pi\sqrt{Q-1}}.
   \end{align}
 \end{remark}
 It is convenient to introduce the notation
 \begin{align} \label{def-ome} 
   \ome \ := \ \frac ts \ 
   \label{le-sc}= \ 2\pi (Q-1) \frac{\eps}{|\ln \eps|}
 \end{align}
 for the aspect ratio of the film.  Combining the previous formulas for the
 stray field we obtain an asymptotic representation of the energy:
 \begin{lemma}[Representation for $E_\eps$] \label{lem-Erep} %
   Let $\Ome \CUS \R^2$ be a set of finite perimeter. Let $M = (M',M_3) \in \AA$
   and let $\OL M = m \otimes \chi_{[0,1]}$ be given by \eqref{def-avg}. Then
   \begin{align} \label{E-FG} %
     E_\eps[M] \
     &=  \ F_\eps[m]  + G_\eps[M]  \\
     &\qquad + \OO\Big(\frac{\eps}{|\ln \eps|} \int_{\Ome_1} |\nabla' M|^2 +
       \frac{|\ln \eps|}{\eps}\int_{\Ome_1}|\p_3 M|^2 + \frac{\alp_0|\ln
       \eps|}{\eps} |B_{\ome}(\p \Ome) \cap \Ome| \Big).
   \end{align} 
   where $\alp_0 = 1$ if $M \neq \OL M$ and $\alp_0 = 0$ else and where
   (with notations \eqref{def-eps}, \eqref{def-ome})
    \begin{align}
      G_\eps[M] \ 
      &= \ \eps |\ln \eps| \int_{\Ome_1} \Big( |\nabla' M|^2 -
        |\nabla' \OL M|^2 \Big)  + \frac {|\ln \eps|^3}{\eps(Q-1)^2(2\pi)^2}\int_{\Ome_1}|\p_3 M|^2 \\
      &\qquad + \frac 18 \iint_{\Ome \times \Ome}  \Big(1- \Gam(\tfrac 1\ome |x-x'|) \Big)\frac{|m_3(x) - m_3(x')|^2}{|x-x'|^3} \ dx d x'   \\
      &\qquad + \frac 14 \iint_{\Ome^c \times \Ome} \Gam(\tfrac 1\ome |x-x'|) \frac{1 -
        m_3(x')^2}{|x-x'|^3} \ dx d x' \\
      &\qquad + \frac 14 \iint_{\Ome \times \Ome}  \Theta(\tfrac 1\ome |x-x'|)  \DIV m'(x) \DIV m'(x') \ dx dx' \\
      &\geq \ 0. 
    \end{align}
\end{lemma}
\begin{proof}
  Let $\t \Ome = s \Ome$, $\t \Ome_t := \t \Ome \times [0,t]$ and
  $\t M(sx,tx_3) := M(x,t)$. By Theorem \ref{thm-strayred} and Theorem
  \ref{thm-stray-rep} we have
  \begin{align}
    \int_{ \R^3} |\HH[\t M]|^2 \ 
    &= \ \int_{ \R^3} |\HH[\OL{\t M_3}e_3]|^2 + \int_{ \R^3} |\HH[\OL{\t M'}]|^2 + \alp_0 \OO \Big( t |B_{t}(\p \t \Ome) \cap \Ome| + t^2 \int_{\t \Ome_t}|\nabla \t M|^2  \Big).
  \end{align}
  We rewrite the expression on the right hand side using Theorem
  \ref{thm-stray-rep}.  We also rescale length and use the notations from the
  statement of the lemma. We also use the fact that $|m| = \chi_\Ome$.  With
  $\Ome_1 := \Ome \times [0,1]$ we then obtain
  \begin{align}
    \EE[\t M]\ 
    &=  \ \FF[m] + \GG[M] + \OO\Big( t s^2 |B_{\ome}(\p \Ome) \cap \Ome|  + t^3 \int_{\Ome_1}   |\nabla' M|^2 + t s^2  \int_{\Ome_1} |\p_3 M|^2  \Big).
  \end{align}
  where with $(..) := (\tfrac st |x-x'|)$ we have
  \begin{align}
    \FF[m] \ &= \ \int_{\Ome} t d^2  |\nabla m|^2  + s^2 t(Q-1)  (1- m_3^2) \ dx 
               - \frac{s t^2}{8\pi} \iint_{\Ome \times \Ome}  \Gam(..) \frac{|m_3(x) - m_3(x')|^2}{|x-x'|^3} \\
    \GG[M] \ 
             &=  \ t d^2 \Big( \int_{\Ome_1} |\nabla' M|^2 - |\nabla \OL M|^2 \Big) + \frac {s^2 d^2}{t} \int_{\Ome_1} |\p_3 M|^2    \\ 
             &\qquad+ \frac{s t^2}{4\pi} \bigg( \iint_{\Ome^c  \times \Ome}  \Gam(..) \, \frac{1 - m_3(x')^2}{|x-x'|^3}  +  \iint_{\Ome \times \Ome}  \Theta(..)  \, \frac{\DIV m'(x) \DIV m'(x')}{|x-x'|} \bigg).
  \end{align}
  We recall that $E_\eps = \frac \pi{st^2}\EE$ so that every prefactor needs to
  multiplied with $\frac \pi{st^2}$.  Using the identities from Remark
  \ref{rem-parel} we compute, $\frac{\pi d^2}{st} = \frac 12 \eps |\ln \eps|$,
  $\frac{\pi(Q-1)s}{t} = \frac {|\ln\eps|}{2\eps}$ and
  $\frac{s d^2}{t^3} = \frac {|\ln \eps|^3}{\eps(2\pi)^3}$. This yields the
  representation formula.  We finally note that $G_\eps \geq 0$ follows from
  $0 \leq \Gam \leq 1$ and $0 \leq \Theta \leq 1$.
\end{proof}

\section{Lower bound and compactness} \label{sec-lower} %

In this section, we give a lower bound for the energy for both energies $F_\eps$
and $E_\eps$. Throughout the section, we use the notation
$F_\eps[m] = L_\eps[m] - N[m]$ where
\begin{align} \label{def-LNeps} 
  L_\eps[m] \ 
  &:= \ |\ln \eps| \int_\Ome \frac \eps 2 |\nabla m|^2 + \frac 1{2\eps}
    (1-m_3^2)^2 \ dx, \\ 
  N[m] \ 
  &:= \ \frac 18 \int_{\Ome} \int_{\Ome} \frac 1{|x-y|^3}|m_3(x) - m_3(y)|^2 \
    dx dy. 
\end{align}
One important ingredient for the proof of the lower bound is the following
interpolation inequality which estimates the $H^{1/2}$--finite difference norm
in terms of BV and $L^\infty$--norm with a small term of higher order. We note
that the leading order constant $8$ in \eqref{eq-interpolation} below is
sharp. The analogous inequality in the case of the torus has been shown in
\cite{KMN-2019}. We state the lemma in a slightly different form, adapted to the
case of finite domains. For the convenience of the reader we sketch the proof
and refer to \cite{KMN-2019} for details.
\begin{lemma}[Interpolation inequality]\label{lem-interpolation} 
  Let $\Ome$ be convex with $\d := |\diam \Ome|$. Let
  $f \in C^\infty(\OL{\Ome})$. Then for any $0 < r \leq R$ we have
  \begin{align}\label{eq-interpolation}
    \hspace{6ex} & \hspace{-6ex} 
                   \int_{\Ome} \int_{\Ome} \frac 1{|x-y|^3}|f(x) - f(y)|^2  \ dx dy \ 
                   \leq \ \pi r \int_{\Ome}|\nabla f|^2 \ dx \\
                 &+ 8 \ln \frac Rr \NI{f}
                   \int_{\Ome}|\nabla f|dx  +  \frac{4\pi \alp_0 }{R} \NI{f}  \min \Big \{ \d   \int_{\R^2} |\nabla f| \ dx, 2 |\Ome| \NI{f} \Big \},
  \end{align}
  where $\alp_0 = 1$ if $R < \d$ and $\alp_0 = 0$ else.
\end{lemma}
\begin{proof}
  Let $X$ be the left hand side of \eqref{eq-interpolation}.  Without loss of
  generality, we may assume that $f$ is not constant. We show that for
  $0 < r \leq R$ and $\Ome_z := \Ome - z$ we have
  \begin{align}
    &\label{intpol-small} 
      \int_{B_r} \int_{\Ome \cap \Ome_z} 
      \frac{|f(x+z)-f(x)|^2}{|z|^3} \ dx dz  \ 
      \leq \ \pi r\int_{\Ome} |\nabla f|^2 \ dx,\\ 
    &\label{intpol-medium} 
      \int_{B_R \setminus B_r} \int_{\Ome \cap \Ome_z} 
      \frac{|f(x+z)-f(x)|^2}{|z|^3} dx dz \ 
      \leq \ 8 \ln \Big(\frac Rr \Big) \NI{f} \int_{\Ome} |\nabla f| \ dx, \\
    &\label{intpol-large} 
      \int_{B_R^c} \int_{\Ome \cap \Ome_z}   \frac{|f(x+z)-f(x)|^2}{|z|^3} dx dz 
      \leq \frac{4\pi}{R} \NI{f}  \min \Big \{ \d   \int_{\R^2} |\nabla f|, 2 |\Ome| \NI{f}  \Big \}. 
  \end{align}
  The assertion of the lemma then follows by adding \eqref{intpol-small} --
  \eqref{intpol-large}. The proof of these estimates follows similarly as in
  \cite{KMN-2019}.  For the convenience of the reader we sketch the details: We
  note that for convex sets $\Ome$ and for $p \geq 1$ we have
  \begin{align}\label{eq-Lptranslation}
      \int_{\Ome \cap \Ome_z} |f(x+z) - f(x)|^p \ dx \ 
      \leq \ \int_{\Ome}|\nabla f(x) \cdot z|^p \ dx \qquad \text{$\forall z \in \R^2$}.
  \end{align}
  Indeed, by the Fundamental Theorem of Calculus and Jensen's inequality,
  \begin{align}
    \hspace{6ex} & \hspace{-6ex} 
                   \int_{\Ome \cap \Ome_z} |f(x+z) - f(x)|^p \ dx \ 
      = \ \int_{\Ome \cap \Ome_z} \int_0^1 \Big| \nabla f(x+tz) \cdot z \ dt|^p \ dx \\ 
      &= \ \int_0^1 \int_{\Ome \cap \Ome_z}  |\nabla f(x+tz) \cdot z|^p \  \ dx dt \ 
      \leq \ \int_{\Ome}|\nabla f(x) \cdot z|^p \ dx \qquad \text{$\forall z \in \R^2$}, 
  \end{align}
  noting that $x + [0,1]z \in \Ome$ for $x \in \Ome \cap \Ome_z$ since $\Ome$ is
  convex. The estimates \eqref{intpol-small} and \eqref{intpol-medium} then
  follow analogously as in \cite{KMN-2019}.  \DETAILS{ For the proof of
    \eqref{intpol-small} we note that by \eqref{eq-Lptranslation} with $p=2$ and
    since $\Ome \cap \Ome_z \SUS \Ome$ we have
    \begin{align}\label{intpol-small_p1}
      \hspace{6ex} & \hspace{-6ex} 
                     \int_{B_r} \int_{\Ome \cap \Ome_z}  \frac{|f(x+z) - f(x)|^2}{|z|^3} dz dx \ 
                     \lupref{eq-Lptranslation}\leq \ \int_{B_r} \int_{\Ome} \frac{|\nabla f(x)\cdot z|^2}{|z|^3} \ dx dz \\
                   &= \ \left(\int_{0}^r
                     \int_{0}^{2\pi} \cos^2 \phi \, d\phi d\rho
                     \right) \left(\int_{\Ome}|\nabla f(x)|^2
                      dx\right)  \ 
                     = \ \pi r\int_{\Ome}|\nabla f|^2dx.
    \end{align}
    For the proof of \eqref{intpol-medium} we calculate
    \begin{align} \label{eq-int2_a} 
      \int_{B_R \setminus B_r} \int_{\Ome_z} \frac{|f(x+z) - f(x)|^2}{|z|^3} \
      dz dx 
      \lupref{eq-Lptranslation}\leq \ 2\|f\|_\infty \int_{\Ome} \int_{B_R
      \setminus B_r} \frac{|\nabla f(x) \cdot z|}{|z|^3} dz dx.
    \end{align}
    The estimate then follows together with the identity
    \begin{align}  
      \int_{B_R \setminus B_r} \frac{|\nabla f(x) \cdot
      z|}{|z|^3}dz \ 
      = \ \int_r^R \int_0^{2\pi} \frac{|\nabla  f(x)|\ |\cos \phi|}{\rho} d\phi d\rho \ 
      = \ 4 \ln\left(\frac{R}{r}\right)|\nabla f(x)|.
    \end{align}
    }
    For the proof of \eqref{intpol-large} we estimate
    \begin{align}
      \int_{B_R^c} \int_{\Ome \cap \Ome_z}   \frac{|f(x+z)-f(x)|^2}{|z|^3} dx dz \ 
      \leq \  2\NI{f}  \int_{B_R^c} \frac 1{|z|^3} \int_{\Ome \cap \Ome_z}   |f(x+z)-f(x)| dx dz. 
    \end{align}
    The estimate then follows by noting that \eqref{eq-Lptranslation}
    with $p=1$ entails
    \begin{align}\label{eq-poincare}
      \int_{\Ome \cap \Ome_z}|f(x + z) - f(x)| dx \ 
      \leq \ \min\left\{\d \int_{\R^2} |\nabla f \cdot z|  dx, 2 |\Ome| \NI{f}  \right\}
    \end{align}
    and integrating in $z$.
  \end{proof}
  We turn to the proof of Theorem \ref{thm-critical1} and Theorem
  \ref{thm-critical2}. We first consider the case of a domain with small
  diameter:
  \begin{lemma}[Small diameter] \label{lem-small} %
    Let $\Ome \CUS \R^2$ be convex. Then there is $\eps_0 > 0$ such that for
    $\eps \in (0,\eps_0)$ the following holds: If the diameter of the set $\Ome$
    is bounded by
      \begin{align} \label{diam-bound2} %
        \diam \Ome \ %
        < \ \frac {\pi}{2e} \Big(1 - \frac 2{|\ln \eps|} \Big)
      \end{align}
      then the following holds
    \begin{enumerate}
    \item $\DS \inf_{M \in \AA, M = \OL M} E_\eps[M] \ = \ \inf_{m \in \OL \AA} F_\eps[m] \ = \ 0$. 
    \item $\DS - c |\p \Ome| \ \leq \ \inf_{M \in \AA} E_\eps[M] \ \leq \ 0$.
    \end{enumerate}
    Furthermore, the infimum in (i) is only achieved by either one of the
    configurations $\pm e_3 \chi_\Ome \in \OL \AA$ for $F_\eps$ and
    $(\pm e_3 \chi_\Ome) \otimes \chi_{[0,t]}$ for $E_\eps$.
  \end{lemma}
  \begin{proof}
    The upper bound for the ground state energy in (i) is given by either one of
    the two configurations $\pm e_3 \chi_{\Ome}$ for the energy $F_\eps$ or
    $\pm e_3 \chi_{\Ome_1}$ for the energy $E_\eps$. It hence remains to give a
    proof for the lower bounds. We write $m(x) := \OL M(x,\frac 12)$.

    \medskip

    We first note that by convexity of the set $\Ome$ we have
    \begin{align}
      \frac{|\ln \eps|}\eps |B_\ome(\p \Ome) \cap \Ome| \ %
      \leq \ \frac{|\ln \eps| \ome}\eps |\p \Ome| \ %
      \lupref{def-ome}\leq \  2\pi (Q-1) |\p \Ome| \ = \ c |\p \Ome|.
    \end{align}
    With the estimate
    $\frac \eps 2 |\nabla m|^2 + \frac 1{2\eps} (1-m_3^2) \geq |\nabla m_3|$ we
    get
    \begin{align} \label{eq-Lsmall} %
      L_\eps[m] \ %
      &= \ |\ln \eps| \big( \frac \eps 2 \NTL{\nabla m_3}{\Ome}^2 + \frac
        1{2\eps} \NPL{1-m_3^2}{1}{\Ome}^2 \big) \\ %
      &\geq \ \frac \eps 2 \NTL{\nabla m_3}{\Ome}^2 + (|\ln \eps| - 1)
        \NNN{\nabla m_3}{\Ome}.
    \end{align}
    By an application of Lemma \ref{lem-Erep} and for $\eps < \eps_0$
    sufficiently small we hence get
    \begin{align} \label{ELN-small} %
      E_\eps[M] + c \gam_0 |\p \Ome| \ %
      &\geq \ F_\eps[m] - \frac{\eps}{|\ln \eps|} \NTL{\nabla m}{\Ome}^2 \\
      &\upref{eq-Lsmall}\geq \ \Big( \frac \eps 2 - \frac{\eps}{|\ln \eps|} \Big) \NTL{\nabla m_3}{\Ome}^2 + (|\ln \eps| - 1) \NNN{\nabla m_3}{\Ome}  - N[m]
    \end{align} %
    with $\gam_0 = 1$ if $M \neq \OL M$ and $\gam_0 = 0$ else. We then apply Lemma \ref{lem-interpolation} with
    \begin{align}
      r \ := \ \frac{1}{\pi} \Big( \frac \eps 2 - \frac \eps{|\ln \eps|}  \Big), \qquad\qquad %
      R \ := \ \max \big \{ r, \d \big \}.
    \end{align}
    which yields 
      \begin{align}\label{eq-Nsmall}
        N[m] \ %
        &\leq \ \Big( \frac \eps 2  - \frac \eps{|\ln \eps|} \Big)\NTL{\nabla m_3}{\Ome}^2 + \Big( |\ln \eps| - \ln \Big(\frac {\pi}{2\d} \big(1 - \frac 2{|\ln \eps|} \big)\Big)  \Big)  \NNN{\nabla m_3}{\Ome}.
      \end{align}
      By inserting \eqref{eq-Nsmall} into \eqref{ELN-small} we obtain the estimate
      \begin{align}
        \min \Big\{ E_\eps[M] + \gam_0 c |\p \Ome|, F_\eps[m] \Big \} \ %
         &= \ \ln \Big(\frac {\pi}{2\d e} \big(1 - \frac 2{|\ln \eps|} \big)\Big)   \NNN{\nabla m_3}{\Ome} \
          > \ 0.
      \end{align}
      which yields both assertions (i) and (ii).
    \end{proof}
    We next give the proof for a general lower bound for the energy which is
    relevant for larger sets and some corresponding estimates for low energy
    states:
    \begin{proposition}[Lower bound] \label{prp-lowfull} %
      Let $\Ome \CUS \R^2$ be convex and let $\eps \in (0, \frac 14)$. Then
    \begin{align} \label{F-thelob-4} %
      \min \Big \{ \inf_{M \in \AA} E_\eps[M] + c |\p \Ome|, \inf_{m \in \OL
      \AA} F_\eps[m] \Big \} \ %
      \geq \ - \frac{\pi^2 e }4 |\Ome|. %
    \end{align}
    Furthermore, there is $C > 0$ such that
    \begin{align} 
      \Big|\ln \Big(\frac{4\NNN{\nabla \OL M_3}{\Ome_1}}{\pi^2 e^2|\Ome|} \Big) \Big| \NNN{\nabla \OL M_3}{\Ome_1}   \ %
      &\leq \ E_\eps[M] +  \frac{\pi^2 e }4  |\Ome| + c |\p \Ome|, \\
      \Big|\ln \Big(\frac{4\NNN{\nabla m_3}{\Ome_1}}{\pi^2 e^2|\Ome|} \Big) \Big| \NNN{\nabla m_3}{\Ome_1}   \ %
      &\leq \ F_\eps[m] + \frac{\pi^2 e }4 |\Ome|. \label{X-F-est-2}
    \end{align}
    If the configuration $M \in \AA$ satisfies for some $\alp > 0$
    \begin{align}
      E_\eps[M] \ \leq \ - \alp |\Ome|, %
    \end{align}
    then for constants $C_\alp, c_\alp$ depending on $\alp$ (with Notation
    \eqref{def-LNeps}) we have
  \begin{enumerate}
  \item\label{it-Xbound} %
    $c_\alp |\Ome| \ \leq \ \NNN{\nabla \OL M}{\Ome_1} \ \ %
    \leq \ C_\alp |\Ome|$,
  \item\label{it-XYnear}
    $\DS 0 \ \leq \ \Big( \frac \eps2 \NTL{\nabla \OL M}{\Ome_1}^2 + \frac 1{2\eps} \NPL{1 -
      \OL M_3^2}{1}{\Ome_1} \Big) - \NNN{\nabla \OL M}{\Ome_1} \ %
    \leq \ \frac{C_\alp}{|\ln \eps|} |\Ome|$,
  \item\label{it-LNest} %
    $\DS c_\alp |\Ome| |\ln \eps| \ %
    \leq \ L_\eps[\OL M],N[\OL M] \ %
    \leq \ C_\alp |\Ome| |\ln \eps|$.
  \item\label{it-N3est} %
    $\DS \NTL{\p_3 M_3}{\Ome_1}^2 \ %
    \DS \leq \ \frac{\eps}{|\ln \eps|^3} \Big( E_\eps[\OL M] + C \Big)$.
  \end{enumerate}
  If $m \in \OL \AA$ satisfies $F_\eps[m] \leq -\alp |\Ome|$, then
  \ref{it-Xbound}--\ref{it-LNest} hold with $\OL M$ replaced by $m$.
  \end{proposition}
  \begin{proof}
    We write
    $m = \OL M(x,\frac 12)$.

    \medskip
    
    \textit{Proof of \eqref{F-thelob-4} and \eqref{X-F-est-2}:} As in the proof
    of Lemma \ref{lem-small} for $\eps < \eps_0$ sufficiently small and in view
    of $F_\eps = L_\eps - N$ we have \eqref{ELN-small}--\eqref{eq-Lsmall}, i.e.
  \begin{align} \label{ELN-small-2} %
                   E_\eps[M] + c |\p \Ome| \
    &
                                               \geq \ F_\eps[m] - \frac{\eps}{|\ln \eps|} \NTL{\nabla  m}{\Ome}^2 \\ %
                 &\upref{eq-Lsmall}\geq \ \Big(1 - \frac {2}{|\ln \eps|}\Big)
                   \frac \eps 2 \NTL{\nabla m}{\Ome}^2 + \big( |\ln \eps| - 1
                   \big) \NNN{\nabla m}{\Ome} - N[m].
  \end{align}
  Applying Lemma \ref{lem-interpolation} with $f = m_3$ and with the choice
  \begin{align}
    r \ := \ \frac{\eps}{2\pi} \Big( 1 - \frac 2{|\ln \eps|}  \Big), \qquad\qquad %
    R \ := \ \frac{\pi |\Ome|}{\NNN{\nabla m}{\Ome}}
  \end{align}
  we get (also using that $\ln y + 1 = \ln (e y)$)
  \begin{align}\label{Nest-2}
    N[m] \ %
    &\leq \  \Big( 1 - \frac 2{|\ln \eps|}  \Big) \frac{\eps}{2}  \NTL{\nabla m}{\Ome}^2
      + \big(\ln \frac{2\pi R}{\eps (1 - \frac {2\eps}{|\ln \eps|})} \big) \NNN{\nabla m}{\Ome} + %
      \frac{\pi |\Ome|}{R} \\ %
    &= \ \Big( 1 - \frac 2{|\ln \eps|}  \Big) \frac{\eps}{2}  \NTL{\nabla m}{\Ome}^2 +  \Big( |\ln \eps|  +\ln \frac{2 e \pi^2 |\Ome|}{1 - \frac {2\eps}{|\ln \eps|}}  \Big) \NNN{\nabla m}{\Ome}.
  \end{align}
  Inserting \eqref{Nest-2} into \eqref{ELN-small-2} (again using that
  $\ln y + 1 = \ln (e y)$) we arrive at
  \begin{align}\label{eq-F1}
    \min \big \{ E_\eps[M] + c |\p \Ome|,   F_\eps[m] \big \} \ %
      \geq \  -
      \DS \ln \Big( \frac{\pi^2 e^2 |\Ome|}{4\NNN{\nabla m}{\Ome}} \Big) \NNN{\nabla m}{\Ome}. %
  \end{align}
  As we do not know the value of $\NNN{\nabla m}{\Ome}$, we minimize the right
  hand side of \eqref{eq-F1} over all values $X := \NNN{\nabla m}{\Ome} \geq 0$:
  This yields the lower bound $- \frac{\pi^2 e }4 |\Ome|$ and hence
  \eqref{F-thelob-4}. Estimate \eqref{X-F-est-2} then follows from \eqref{eq-F1}
  as well.

  \medskip

  \textit{Proof of \ref{it-Xbound}--\ref{it-N3est}:} Estimate \ref{it-Xbound}
  follows from \eqref{X-F-est-2}.  In the estimate \eqref{ELN-small-2} we have
  neglected the nonnegative term
  \begin{align}
    (|\ln \eps| - 1)  \Big( \int_{\Ome} \frac \eps 2|\nabla m_3|^2 + \frac 1{2\eps} (1-m_3^2) \ dx  -  \int_{\Ome}|\nabla m_3| \ dx  \Big). 
  \end{align}
  Keeping this term yields additionally the estimate \ref{it-XYnear}. Estimate
  \ref{it-LNest} follows from \ref{it-Xbound} and \ref{it-XYnear} and the
  definition \eqref{def-LNeps}
\end{proof}
The proof of the compactness statement in Theorem
\ref{thm-critical2}\ref{it-compact} is a straightforward consequence of the
BV--bound of Lemma \ref{prp-lowfull}:
\begin{proposition}[Compactness] \label{prp-compact} \text{}%
  Let $\Ome \SUS \R^2$ be convex.
  \begin{enumerate}
  \item For any sequence $m_\eps \in \OL \AA$ with
    $\limsup_{\eps \to 0} F_\eps[m_\eps] < \infty$ there is
    $m \in BV(\Ome, \pm e_1)$ with $|m| = \chi_{\Ome}$ and a subsequence with
    $m_\eps \to m$ \text{in $L^1(\Ome)$} for $\eps \to 0$.
  \item The assertion of Theorem \ref{thm-critical2}\ref{it-compact} holds.
  \end{enumerate}
\end{proposition}
\begin{proof}
  By Lemma \ref{prp-lowfull}\ref{it-Xbound} we have
  $\NNN{\nabla m_{3,\eps}}{\Ome} \ \leq \ C$.  This shows that
  $m_{3,\eps} \to m_3$ in $L^1(\R^2)$ for some $m_{3,\eps} \in L^1(\R^2)$ and
  for a subsequence by the compact embedding of $BV(\Ome)$ into
  $L^1(\Ome)$. From Proposition \ref{prp-lowfull}\ref{it-LNest} we also get
  \begin{align}
    \int_\Ome m_{1,\eps}^2 + m_{2,\eps}^2 \ dx \ %
    = \ \int_\Ome  (1 - m_{3,\eps}^2) \ dx \ %
    \leq \ \eps L_\eps[m_\eps] \ %
    \leq \ C \eps |\ln \eps| \ %
    \to \ 0.
  \end{align}
  Altogether this implies the assertion (i). Similarly, the corresponding
  compactness holds for $\OL M$ which again follows from Proposition
  \ref{prp-lowfull}. This shows (ii).
\end{proof}

\section{Proof of upper bound} \label{sec-upper} %

We recall the family of asymptotically one--dimensional profiles from
\cite{KMN-2019}:
\begin{lemma}[{\cite[Lem. 4.2]{KMN-2019}}]\label{lem-opt}
  For $0 < \eps < 1$, let $\xi_{\eps} \in C^1(\R)$ be given by
    \begin{align}\label{eq-xi_eps_R}
      \xi_{\eps}(\rho) \ %
      := \ \sin \left( \frac \pi 2
      \frac{\arcsin ( \tanh ( \rho /
      \eps ) )}{\arcsin ( \tanh (\eps^{-\frac 12}) )} \right) \qquad |\rho| < \eps^{\frac 12}
    \end{align}
    and $\xi_\eps(\rho) = \mathrm{sign}(\rho)$ for
    $|\rho| \geq \eps^{\frac 12}$.  Then for universal $C,c_1,a > 0$, we have
    \begin{enumerate}
    \item \label{eq-transition-local}
      $\DS \frac12 \int_\R \Big( 
      \frac{\eps |\xi_\eps'|^2}{1-\xi_\eps^2} +
      \frac{1-\xi_\eps^2}{\eps} \Big) d\rho \ %
      \leq \ 2 + C e^{-a \eps^{-1/2}}$,
    \item\label{it-transition-nonlocal}  %
      $\DS \frac 14 \int_{-H}^{H} \int_{-H}^H
      \frac{|\xi_\eps(\rho)- \xi_\eps(\rho')|^2}{|\rho-\rho'|^2}
      d\rho d\rho' \geq \ 2 \ln \left( \frac{c_1 H}{\eps} \right) \qquad \text{for }H %
              \geq 2\eps$,
            \item\label{it-transition-nonlocal-lot} %
              $\DS \frac 14 \int_{-H}^{H} \int_{-H}^H \frac{|\xi_\eps(\rho)-
                \xi_\eps(\rho')|^2}{|\rho-\rho'|} d\rho d\rho' \leq \ C H$,
            \item\label{it-transition-nonlocal-lot-2} %
              $\DS \frac 14 \int_{-H}^{H} \int_{-H}^H |\xi_\eps(\rho)-
                \xi_\eps(\rho')|^2 \ d\rho d\rho' \leq \ C H^2$.
    \end{enumerate}
\end{lemma}
\begin{proof}
  (i) and (ii) follow from {\cite[Lem. 4.2]{KMN-2019}} with the special choise
  of a transition layer with support of size $2\eps^{\frac 12}$. (iii) and (iv)
  are straightforward estimates.
\end{proof}
For the special case $\lambda = 0$, the $\Gam$-convergence and in particular the
construction of a recovery sequence is a classical result, relying on the
optimal one-dimensional transition profiles to smooth out the jump discontinuity
in the limit configuration \cite{ABV}.  As it turns out, this construction also
works for $\lambda >0$, where $F_{\eps, \lambda}$ is nonlocal. We will use a
similar construction based on the nearly optimal profile $\xi_\eps$ from Lemma
\ref{lem-opt}. As the calculations for the local part of the energy are
well-known, our focus is on the contribution of the homogeneous
$H^{1/2}$-norm. We define the finite range version of the energy $F_\eps$ by
\begin{align} \label{def-tF} 
  F_{\eps,R}[m] &= |\ln \eps| \int_{\Ome} \frac{\eps}2 |\nabla m|^2+\frac
              1{2\eps} (1-m_3^2) \ dx - \frac 18 \iint_{\R^2, |x-y| \leq R} \frac {|m_3(x)
              - m_3(x')|^2}{|x-x'|^3} dx dx'
\end{align}
noting that for all $m \in \OL \AA$ we have
\begin{align}
  F_\eps[m] \ \leq \ F_{\eps,R}[m].
\end{align}
We next give an upper bound construction related to circular domains:
\begin{lemma}[Upper bound construction for disk] \label{lem-disk} %
  Suppose that $B_{2R} \SUS \Ome \SUS \R^2$ for some $R > 0$. There is a
  universal constant $R_0 > 0$ such that if $R > R_0$ then there are sequences
  $m_\eps \in \OL \AA$ such that
  \begin{align} \label{disk-bc} %
    \text{$m_\eps = -e_3$ \ in 
    $B_{R-\eps^{1/2}}$ \qquad and  \
    $m_\eps = -e_3$ in
    $\Ome \BS B_{R + \eps^{1/2}}$ }
  \end{align}
  and for $\eps > 0$ sufficiently small we have
  \begin{align} \label{dis-est} %
    F_{\eps,R}[m_\eps] %
    \ \leq \ - C.
  \end{align}
\end{lemma}
\begin{proof}
  We proceed similarly as in the construction of \cite[Lem. 4.3]{KMN-2019},
  however with a slightly more careful estimate since we need to capture the
  sign of the leading order term of the energy. We write $d(x) := R - |x|$ for the
  signed distance function for $\p B_R$. For
  $x \in \SS_\eps := B_{R+\eps^{1/2}} \BS B_{R-\eps^{1/2}}$ we set
  \begin{align}\label{eq-meps}
    m_{\eps}(x) \ := \ \xi_\eps(d(x)) e_3 + \sqrt{1-\xi_\eps^2(d(x))}
    \ \  \frac{x^\perp}{|x|},
  \end{align}
  where $\xi_\eps$ is given in Lemma \ref{lem-opt}.  We also set $m_\eps = e_3$
  in $B_{R-\eps^{1/2}}$ and $m_\eps = -e_3$ in $\Ome \BS B_{R + \eps^{1/2}}$.
  Then $m_\eps \in \OL \AA$ and satisfies \eqref{disk-bc}.

  \medskip  

  It remains to show \eqref{dis-est}: We use the coarea formula to integrate on
  the level sets of $d$. By Lemma \ref{lem-opt} we then get
    \begin{align}
      L_\eps[m_\eps] \ %
      &:= \ |\ln \eps| \int_{\Ome} \Big(\frac{\eps}{2}|\nabla m_\eps|^2 +
        \frac{1}{2\eps}(1-{m_{\eps,3}^2} ) \Big)dx\notag \\ %
      &= \ |\ln \eps| \int_{-\eps^{\frac 12}}^{\eps^{\frac 12}} \Big(\frac{\eps
        |\xi'_\eps(\rho)|^2}{2(1-\xi_\eps^2(\rho))} +
        \frac{1}{2\eps}(1-\xi_\eps^2(\rho))\Big) \ 2\pi (R + \rho)) \ d\rho \\ %
      &\stackrel{\ref{eq-transition-local}}= \  2 |\p  B_R| |\ln \eps| + \OO(\eps^{1/2}). \label{upper-LL} 
    \end{align}
  For $\ell < \pi R$ and $H < R$ to be chosen later, let
  $Q := (-\ell, \ell) \times (-H,H)$. We define the diffeomorphism
  $\Phi : Q \to \Phi(Q)$ by $\Phi(s,\rho) = (R + \rho) \nu(s)$ with
  $\nu(s) := (\cos \frac{s}{2\pi R}, \sin \frac{s}{2\pi R})$, i.e. $s$ is the
  geodesic distance on $\p U$ from $e_1$ and $\rho$ is the distance to $\p U$.
  By rotational symmetry
    \begin{align} \label{upper-N1} %
      N[m] \ %
      &:= \ \frac 14 \int_{\R^2} \int_{\R^2} \frac{|m_{\eps,3}(x) -
        m_{\eps,3}(y)|^2}{|x-y|^3} \ dx dy \\ %
      &\geq \ \frac {|\p U|}4 \iint_{(-H, H)^2} \bigg( \int_{-\ell}^\ell
        \frac{1}{|\Phi(\rho,0)-\Phi(\rho',s)|^3} \ ds \bigg) \
        |\xi_{\eps}(\rho') - \xi_{\eps}(\rho)|^2 \Big(1+\frac {\rho'} R \Big) \
        d\rho d\rho'.
    \end{align}
    We consider $X := |(\sig,\rho - \rho')|$ and
    $Y := |\Phi(0,\rho)-\Phi(\sig,\rho')|$.  We have
    $\p_s \Phi(s,\rho) = (1+\frac \rho R) \tau(s)$ and
    $\p_\rho \Phi(s,\rho) \ = \ \nu(s)$.  Since $D\Phi(0,0) = \id$ and
    $\NIL{D^2\Phi}{Q} \leq \frac 2R$ for $|\rho| < R$, by Taylor expansion we
    have $|X-Y| \leq \frac{X^2}R$. We also have $|X-Y| \leq \frac 12 X$ and
    hence $Y \sim X$ for $X \leq \frac R2$.  Therefore, for $X \in Q$ we have
    \begin{align} \label{ivo} %
      \int_{-\ell}^\ell \Big|\frac{1}{X^3} - \frac 1{Y^3} \Big| d\sig %
      &= \int_{-\ell}^\ell \Big|\frac{(X-Y) (X^2 + Y^2)}{X^3Y^3} \Big| d\sig %
        \leq \int_{-\ell}^\ell \frac{C}{R X^2} \ d\sig %
        \leq \ \frac{C}{R |\rho - \rho'|} %
    \end{align}
    where we have used the integral identity
    \begin{align}
      \int_{-\ell}^\ell \frac{1}{s^2 + (\rho-\rho')^2} \ ds \ 
      &= \ \frac{2}{\rho - \rho'} \arctan \Big( \frac {\ell}{\rho - \rho'} \Big) \ %
        \leq \ \frac \pi{|\rho - \rho'|}. \label{eq-tang-int-rest}
    \end{align}
    The second estimate follows since $\NI{\arctan} \leq \frac \pi 2$.  By
    standard integration and using Taylor's formula we also have
    \begin{align}
      \int_{-\ell}^\ell \frac{1}{\big(s^2 + (\rho-\rho')^2\big)^{3/2}} \ ds \ 
      &= \ \frac{2}{(\rho -
        \rho')^2} \Big(1  
        + \frac{(\rho - \rho')^2}{\ell^2}\Big)^{-\frac 12} \ %
        \geq \ \frac{2}{(\rho - \rho')^2} - \frac 1{\ell^2}, \label{eq-tang-int-2}
    \end{align}
    \DETAILS{By Taylor expansion we have for some
      $|q| \leq |\rho-\rho'|$ 
      \begin{align}
        \frac 1{(\rho -
        \rho')^2} \Big(1 + \frac{(\rho-\rho')^2}{\ell^2}\Big)^{-\frac 12} \ %
        = \ \frac 1{(\rho -
        \rho')^2} \Big( 1 - \frac 12 \frac{q^2}{\ell^2} \Big) 
        = \ \frac 1{(\rho -
        \rho')^2} - \OO \Big(\frac {1}{\ell^2}\Big).
      \end{align}
    }
    We use the above calculations to estimate the inner integral in
    \eqref{upper-N1}, noting that the integrand is pointwise nonnegative
    everywhere. We arrive at
    \begin{align} \label{upper-N1-b} %
      N[m_\eps] \ %
      &\geq \ \frac {|\p U|}4 \iint_{(-H, H)^2} \bigg( \frac 2{|\rho - \rho'|^2}
        - \frac C{R|\rho - \rho'|} - \frac 1{\ell^2} \bigg) \ |\xi_{\eps}(\rho')
        - \xi_{\eps}(\rho)|^2 \Big(1+\frac {\rho'} R \Big) \ d\rho d\rho' \\
      &= \ \frac {|\p U|}4 \iint_{(-H, H)^2} \bigg( \frac 2{|\rho - \rho'|^2} -
        \frac C{R|\rho - \rho'|} - \frac 1{\ell^2} \bigg) \ |\xi_{\eps}(\rho') -
        \xi_{\eps}(\rho)|^2 \ d\rho d\rho',
    \end{align}
    where the identity in the second line above follows from symmetry. An
    application of the last three estimates in Lemma \ref{lem-opt} then yields
    \begin{align} \label{upper-N} %
      N[m_\eps] \ %
      &\geq \ 2|\p U| |\ln \eps| + 2 |\p U| \Big( \ln \frac{c_1 H}{2} - \frac {C
        H}R - \frac {C H^2}{\ell^2} \Big) \ %
        =: \ 2 |\p U| |\ln \eps| + C_1.
    \end{align}
    For $R$ sufficiently large we can hence find $H$ and $\ell$ with
    $1 \ll H \ll \ell \ll R$ such that $C_1$ is positive. Adding equations
    \eqref{upper-LL} and \eqref{upper-N} yields the asserted inequality. In
    particular, since $H \leq R$ the bound holds for the finite range energy
    $F_{\eps,R}$.
\end{proof}
\begin{proposition}[Upper bound]\label{prp-constr}
  Let $\Ome \CUS \R^2$ and let
  $\Ome_{R} := \{ x \in \Ome : \dist(\t \Ome, \Ome^c) \geq R \}$ for $R >
  0$. Then there is $R > 0$ such that the following holds.  There are sequences
  $m_\eps \in \OL \AA$ and $M_\eps(x) := m_\eps \otimes \chi_{[0,t]} \in \AA$
  such that
    \begin{align}
      \max \Big \{ \limsup_{\eps \to 0} E_\eps[M_\eps], \limsup_{\eps \to 0} F_\eps[m_\eps] \Big \} %
      \ \leq \ - C |\Ome_{R}|.
    \end{align}
\end{proposition}
\begin{proof} %
  \textit{Step 1:} In this step we give the proof for the energy $F_\eps$.  Let
  $R_0$ be the constant from Lemma \ref{lem-disk} and let $R := 10 R_0$. Then
  the family of balls $\UU \:= \{ B_R(x) : x \in \Ome \}$ is an open covering of
  $\Ome^{(R)}$. Since $\Ome$ is compact and by Vitali's covering theorem there
  is a finite subcovering $\{ B_i(x_i) : i = 1, \ldots, N \}$ such that the
  balls $\t B_i := B_i(\frac {x_i}5)$ are pairwise disjoint. Let
  $\hat B_i := B_i(\frac {x_i}{10})$. By construction we have
  $\sum_{i=1}^N |\t B_i| \sim |\Ome^{(R)}|$.  For $x \in \hat B_i$ we define
  $m_\eps := m_\eps^{(i)} : B_i \to \S^1$ as in Lemma \ref{lem-disk}. For
  $x \nin \bigcup_i \hat B_i$ we define $m_\eps := e_3$.

  \medskip

  We next apply Lemma \ref{lem-disk}. Since, the domain constructions on the
  single balls have distance $\geq 2R$ we do not need to take the interaction
  energy between the balls for the finite range energy $F_{\eps, R}$ into
  account, cf. \eqref{def-tF}. We hence get
  \begin{align}
    F_\eps[m_\eps] \ %
    \leq \ \sum_{i=1}^N F_\eps[m_\eps^{(i)}] \ %
    \leq \ - \sum_{i=1}^N C |\t B_i| \ %
    \leq \ - C |\Ome^{(R)}|
  \end{align}
  which is (i). In order to show (ii) we apply Theorem \ref{thm-strayred} and
  Theorem \ref{thm-stray-rep}. For this we note that the error term in Theorem
  \ref{thm-strayred} vanishes since $M_\eps = \OL{M_\eps}$. We also note that by
  construction we have $\DIV' M_\eps' = 0$.  With analogous calculations as in
  the proof of Proposition \ref{prp-lowfull} we hence get
  \DETAILS{With $y := x-x'$
  \begin{align}
    E_\eps \ %
    &=  \ L_\eps  %
     - \frac 12 \iint_{\Ome \times \Ome}  \frac{\Gam(\frac 1\ome |y|)}{|y|^3} |m_3(x) - m_3(x')|^2 %
      + \iint_{\Ome^c  \times \Ome}  \frac{\Gam(\frac 1\ome |y|)}{|y|^3} \big( 1 - m_3(x')^2 \big).  %
  \end{align}}
  \begin{align}
    E_\eps[M_\eps] \ %
    &=  \ F_\eps[m_\eps]  \ %
      + \frac 12 \iint_{\Ome \times \Ome}  \frac{1 - \Gam(\frac 1\ome |x-x'|)}{|x-x'|^3} |m_3(x) - m_3(x')|^2 \ dx dx' \\
    &\qquad + \iint_{\Ome^c  \times \Ome}  \frac{\Gam(\frac 1\ome |x-x'|)}{|x-x'|^3} \big( 1 - m_3(x)^2 \big) \ dx dx' \\
    &=: \ F_\eps[m_\eps] + I_\eps + J_\eps.
  \end{align}
  Since $\ome \to 0$ as $\eps \to 0$ we have
  $1 - \Gam(\frac 1\ome |x-x'|) \to 0$ in $\Ome \times \Ome$ pointwise. Since
  $\NI{1 - \Gam} \leq 1$ by dominated convergence we have $I_\eps \to 0$ as
  $\eps \to 0$. Finally, in our construction we have $m_{3,\eps} = 1$ in
  $\Ome \BS \Ome^{(R)}$. Since $\NI{\Gam} \leq 1$ and $\NI{m_3} \leq 1$ we get
  \begin{align}
    J_\eps \ %
    &\leq \ C_R \int_{\Ome^{(R)}} |1 - m_3(x)^2| \ dx \Big) \ %
    \leq \ C_{R} |\Ome|^{\frac 12} \Big( \int_{\Ome^{(R)}} |1 - m_3|^2 \ dx \Big)^{\frac 12} \\
    &\leq \ \eps C_{R} |\Ome|^{\frac 12} \Big( \int_{\Ome^{(R)}} |1 - m_3|^2 \ dx \Big)^{\frac 12}.
  \end{align}
\end{proof}

 \paragraph{Acknowledgements.}
 The authors gratefully acknowledge support from the DFG under Grant No.
 \#392124319 and under Germany's Excellence Strategy EXC-2181/1 – 390900948.

\paragraph{Data availability statement} All data generated or analysed during
this study are included in the main article file

\small
\bibliographystyle{plain} %
\bibliography{mmbib}

\end{document}

%% file: all-def.tex
\newcommand{\CUS}{\Subset}

  { \end{itemize} }
  { \end{enumerate} }

\newcommand{\OL}[0]{\overline} %
\newcommand{\BS}[0]{\backslash}

\newcommand{\FA}[1]{\ensuremath{\forall #1}}

\newcommand{\HW}[1]{} 

\renewcommand{\t}{\tilde}
\newcommand{\p}{\partial}
\renewcommand{\d}{\delta}

\newcommand{\SUS}{\subset}
\newcommand{\Rn}{\R^n}

\newcounter{pcounter}

\renewcommand{\AA}{{\mathcal A}}

\newcommand{\EE}{{\mathcal E}}
\newcommand{\FF}{{\mathcal F}}
\newcommand{\GG}{{\mathcal G}}
\newcommand{\HH}{{\mathcal H}}

\newcommand{\LL}{{\mathcal L}}

\newcommand{\OO}{{\mathcal O}}

\renewcommand{\SS}{{\mathcal S}}

\newcommand{\UU}{{\mathcal U}}

\newcommand{\R}{\mathbb{R}}

\renewcommand{\S}{\mathbb{S}}

\newcommand{\alp}{\alpha}

\newcommand{\gam}{\gamma}
\newcommand{\eps}{\epsilon}
\newcommand{\zet}{\zeta}

\renewcommand{\phi}{\varphi}
\newcommand{\ome}{\omega}
\newcommand{\sig}{\sigma}

\newcommand{\Gam}{\Gamma}

\newcommand{\Ome}{\Omega}

\newcommand{\nin}{\not\in}

\DeclareMathOperator*{\DIV}{div}

\DeclareMathOperator*{\diam}{diam}

\DeclareMathOperator*{\dist}{dist}

\DeclareMathOperator*{\spt}{spt}

\renewcommand{\iint}{\int\!\!\!\!\int}

\def\Xint#1{\mathchoice
{\XXint\displaystyle\textstyle{#1}}%
{\XXint\textstyle\scriptstyle{#1}}%
{\XXint\scriptstyle\scriptscriptstyle{#1}}%
{\XXint\scriptscriptstyle\scriptscriptstyle{#1}}%
\!\int}
\def\XXint#1#2#3{{\setbox0=\hbox{$#1{#2#3}{\int}$}
\vcenter{\hbox{$#2#3$}}\kern-.5\wd0}}

\def\dashint{\Xint-}

\newcommand{\upref}[2]{\hspace{-0.8ex}\stackrel{\eqref{#1}}{#2}} 

\newcommand{\lupref}[2]{\hspace{0ex} \stackrel{\eqref{#1}}{#2}} 

\usepackage{color}
\definecolor{verylightblue}{rgb}{0.95, 0.95, 0.95}  
\definecolor{lightblue}{rgb}{0.7, 0.7, 1}

\definecolor{eqyellow}{rgb}{0.9375,0.8984,0.5469}
\definecolor{subeqyellow}{rgb}{1,0.9373,0.8353}

\definecolor{darkcyan}{rgb}{0.0, 0.45, 0.95} 

\definecolor{mygreen}{rgb}{0.3, 0.6, 0.3} 
\definecolor{verylightgreen}{rgb}{0.95, 0.95, 0.95} 
\definecolor{verydarkgreen}{rgb}{0, 0.5, 0}
\definecolor{darkgreen}{rgb}{0.85, 0.85, 0.85}  
\definecolor{mydarkgreen}{rgb}{0, 0.5, 0} 

\definecolor{mybrown}{rgb}{0.85, 0.4, 0.3}
\definecolor{verylightbrown}{rgb}{0.98, 0.72, 0.58}
\definecolor{verydarkbrown}{rgb}{0.44, 0.26, 0.26}

\definecolor{orange}{rgb}{1, 0.5, 0}

\definecolor{BurntOrange}{rgb}{1,0.356,0}

\definecolor{mydarkred}{rgb}{1,0.086,0.255}

\definecolor{RoseVYDP}{rgb}{0.84,0.086,0.255}

\definecolor{dgreen}{rgb}{0, 0.8, 0.5}     

\definecolor{CanaryBRT}{rgb}{1,0.76,0.26}

\definecolor{cyan}{rgb}{0, 1, 1}
\definecolor{verylightgray}{rgb}{0.95, 0.95, 0.95}
\definecolor{lightgray}{rgb}{0.8, 0.8, 0.8}
\definecolor{verylightred}{rgb}{1, 0.8, 0.78}
\definecolor{verylightyellow}{rgb}{0.99, 0.98, 0.5}

\newcommand{\mygreen}{\color{mygreen}}

\newcommand{\ignore}[1]{{}}

\newcommand{\NNN}[2]{\|#1\|_{#2}}

\newcommand{\NTL}[2]{\|#1\|_{L^2({#2})}}
\newcommand{\NI}[1]{\|#1\|_{L^\infty}}
\newcommand{\NIL}[2]{\|#1\|_{L^\infty({#2})}}

\newcommand{\NPL}[3]{\|#1\|_{L^{#2}(#3)}}

\makeatletter
\newcommand{\VEC}[2][r]{
  \gdef\@VORNE{1}
  \left(\hskip-\arraycolsep%
    \begin{array}{#1}\vekSp@lten{#2}\end{array}%
  \hskip-\arraycolsep\right)}
\def\vekSp@lten#1{\xvekSp@lten#1;vekL@stLine;}
\def\vekL@stLine{vekL@stLine}
\def\xvekSp@lten#1;{\def\temp{#1}%
  \ifx\temp\vekL@stLine
  \else
    \ifnum\@VORNE=1\gdef\@VORNE{0}
    \else\@arraycr\fi%
    #1%
    \expandafter\xvekSp@lten
  \fi}
\makeatother

